\pdfoutput=1
\documentclass[journal=acscii,layout=twocolumn, manuscript=article]{achemso}

\usepackage[english]{babel}
\usepackage[utf8]{inputenc}
\usepackage[colorinlistoftodos, color=green!40, prependcaption]{todonotes}
\usepackage{hyperref}
\addto{\captionsenglish}{} 

\usepackage{amsthm}
\usepackage{mathtools}
\usepackage{physics}
\usepackage{xcolor}
\usepackage{graphicx}
\usepackage{graphics}
\usepackage[fontsize=10]{fontsize}
\usepackage{amsmath}
\usepackage{hyperref}

\newcommand{\mathspace}{\hspace*{0.07cm}}

 \SectionNumbersOn
 \usepackage{caption}
 
\author{Simon Axelrod}
    \affiliation{Department of Chemistry and Chemical Biology,
    Harvard University, Cambridge, MA, 02138}
    \alsoaffiliation{Department of Materials Science and Engineering, Massachusetts Institute of Technology, Cambridge, MA, 02139}

\author{Eugene Shakhnovich}
    \affiliation{Department of Chemistry and Chemical Biology, 
    Harvard University, Cambridge, MA, 02138}
  
\author{Rafael Gómez-Bombarelli}
    \email{rafagb@mit.edu}
    \affiliation{Department of Materials Science and Engineering, Massachusetts Institute of Technology, Cambridge, MA, 02139}

\date{\today} 

\newcommand{\PaperTitle}{Mapping the space of photoswitchable ligands and photodruggable proteins with computational modeling}
\title[]{\PaperTitle}

\begin{document}

\begin{abstract}

Light-activated drugs are a promising way to localize biological activity and minimize side effects. However, their development is complicated by the numerous photophysical and biological properties that must be simultaneously optimized. To accelerate the design of photoactive drugs, we describe a procedure that combines ligand-protein docking with chemical property prediction based on machine learning (ML). We apply this procedure to 58 proteins and 9,000 photo-drug candidates based on azobenzene \textit{cis}--\textit{trans} isomerism. We find that most proteins display a preference for \textit{trans} isomers over \textit{cis}, and that the binding affinities of nominally active/inactive pairs are in fact highly correlated. These findings have significant value for photopharmacology research, and reinforce the need for virtual screening to identify compounds with rare desirable properties. Further, we combine our procedure with quantum chemical validation to identify promising candidates for the photoactive inhibition of PARP1, an enzyme that is over-expressed in cancer cells. The top compounds are predicted to have long-lived active forms, differential bioactivity, and absorption in the near-infrared therapeutic window. 

\end{abstract}
\maketitle

\section*{Introduction}
Photopharmacology is an emerging field that uses light to control the activity of drugs \cite{lerch2016emerging, broichhagen2015roadmap}. The goal of light-controlled bioactivity is to minimize side effects. By activating the drug only at specific locations or times, one can minimize off-target effects, thereby improving quality of life and increasing the maximum deliverable dose \cite{lerch2016emerging}. Such therapeutics have been hypothesized as a path for the treatment of cancer \cite{szymanski2015light},  neurodegenerative diseases \cite{broichhagen2014optical}, bacterial infections \cite{velema2013optical}, diabetes \cite{broichhagen2014optical2}, and blindness \cite{bonardi2010light}.

Photoactive drugs are built around photoswitches, which are molecules that change their properties in response to light. A popular choice is azobenzene \cite{broichhagen2015roadmap}, which exhibits \textit{trans}  $\leftrightarrow$ \textit{cis} isomerism in response to light. Since bioactivity depends strongly on the shape of the ligand, the large structural change upon illumination of azobenzene can change the effect of the drug.

Photoactive drug development is complicated by the number of properties that must be optimized. In addition to regular drug properties, photoactive compounds must absorb light in a narrow range of the near-infrared (IR), since human tissue is only transparent to light in this region \cite{dong2017near}. They must also have a high density of the active isomer under steady-state illumination, and display differential bioactivity between the two isomers. Lastly, they must thermally revert to the stable isomer in a specific time frame. For photoactive drugs one typically wants the unstable isomer to be active for as long as possible, while for ion channel blockers, the target lifetime is usually milliseconds \cite{mourot2011tuning}. Optimizing these properties through trial-and-error experimentation is both slow and costly \cite{axelrod2022learning}.

To accelerate the development of photoactive drugs, we \cite{axelrod2022excited, axelrod2022thermal} and others \cite{adrion2021benchmarking, mukadum2021efficient, griffiths2022data} have developed computational tools to predict the photophysical properties of azobenzene derivatives. In Ref. \cite{axelrod2022excited} we developed a machine learning (ML) force field to predict the quantum yield and absorption spectrum of photoswitches. In Ref. \cite{axelrod2022thermal} we refined this force field and developed an automated workflow to predict the thermal half-lives of \textit{cis} isomers. 

In this work, we combine our models with computational docking to predict the chemical properties and bioactivity of photoactive drug candidates across a broad space of ligands and protein targets. We first use docking to predict the \textit{cis} and \textit{trans} binding free energy of 9,000 ligands to 58 different proteins. The results demonstrate the extent to which \textit{cis} or \textit{trans} isomers are systematically favored to bind a given target, and the correlation between \textit{cis} and \textit{trans} docking scores. The scores and poses are available  at \url{https://doi.org/10.18126/41tz-igf4} through the Materials Data Facility \cite{blaiszik2016materials, blaiszik2019data}. We then combine these results with photophysical properties to virtually screen the 9,000 compounds for photoactive inhibition of the PARP1 cancer target. We identify three hits with predicted differential activity, redshifted absorption, and long thermal half-lives. The photophysical properties are validated for two candidates with quantum chemistry. Full molecular dynamics (MD) simulations of the absorption spectra suggest that both absorb in the near-IR. Lastly, we show that both docking and ML-accelerated simulations can be further sped up with graph-to-property ML models. This paves the way for rapid virtual screening of large photoactive drug libraries.

\begin{figure*}[t!]
    \centering
    \includegraphics[width=\textwidth]{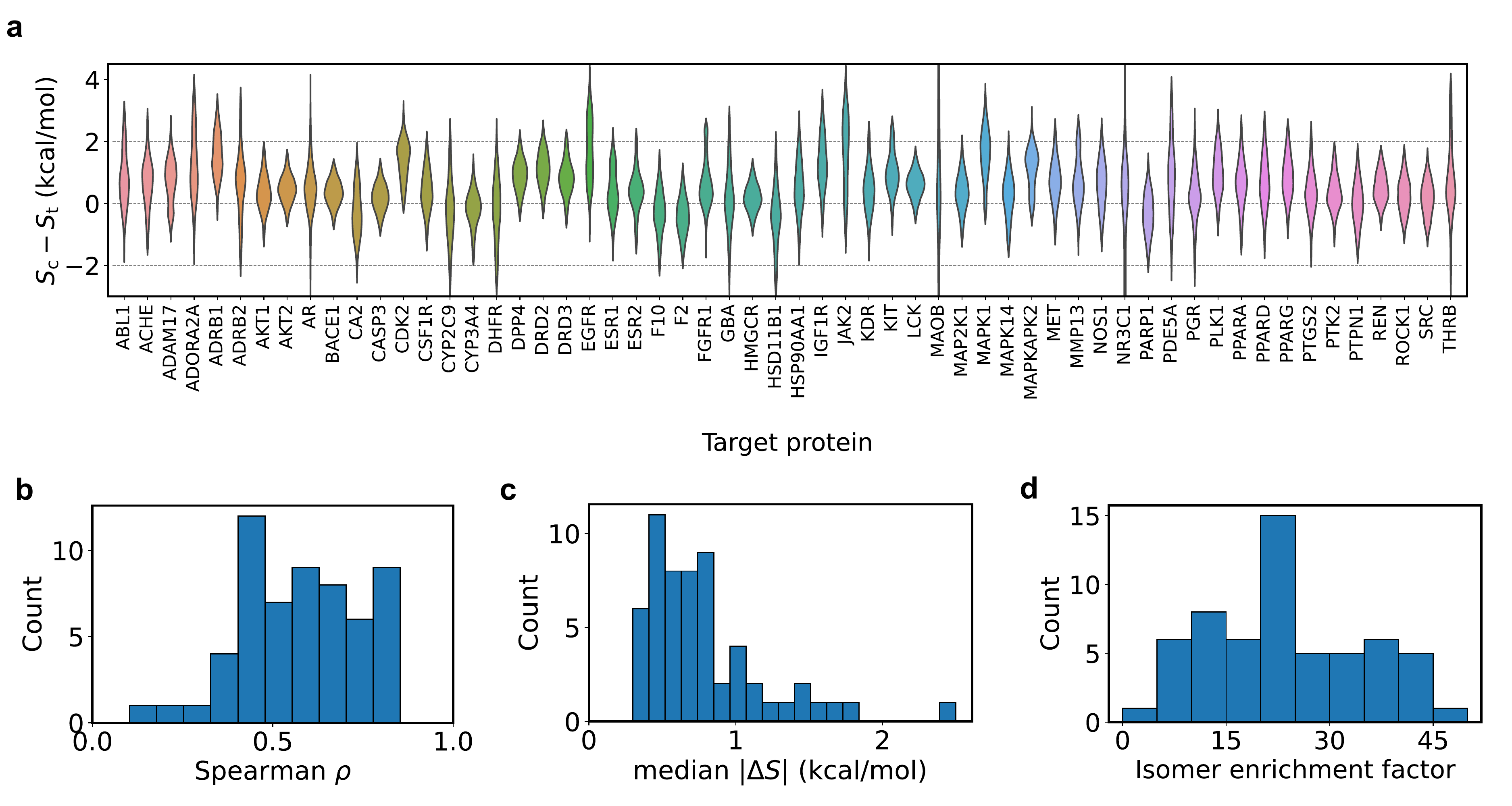}
    \caption{Comparison of docking scores for the \textit{cis} and \textit{trans} isomers of each ligand. (a) Distribution of \textit{cis} score minus \textit{trans} score for different targets. 
    Lines for 0 and $\pm 2$ kcal/mol are shown in dashed gray. (b) Distribution of Spearman rank coefficients between \textit{cis} and \textit{trans} docking scores for each target. (c) Median absolute score difference between isomers for each target. (d) Enrichment factor of \textit{cis} hits using scores from the associated \textit{trans} ligands. This measures the likelihood of a \textit{cis} compound being a hit if the associated \textit{trans} isomer is also a hit. Hits are defined as ligands in the top 1\% of all docking scores. The distribution of enrichment factors is shown for different targets.}
    \label{fig:cis_trans_correlation}
\end{figure*}

\begin{figure*}[t!]
    \centering
    \includegraphics[width=\textwidth]{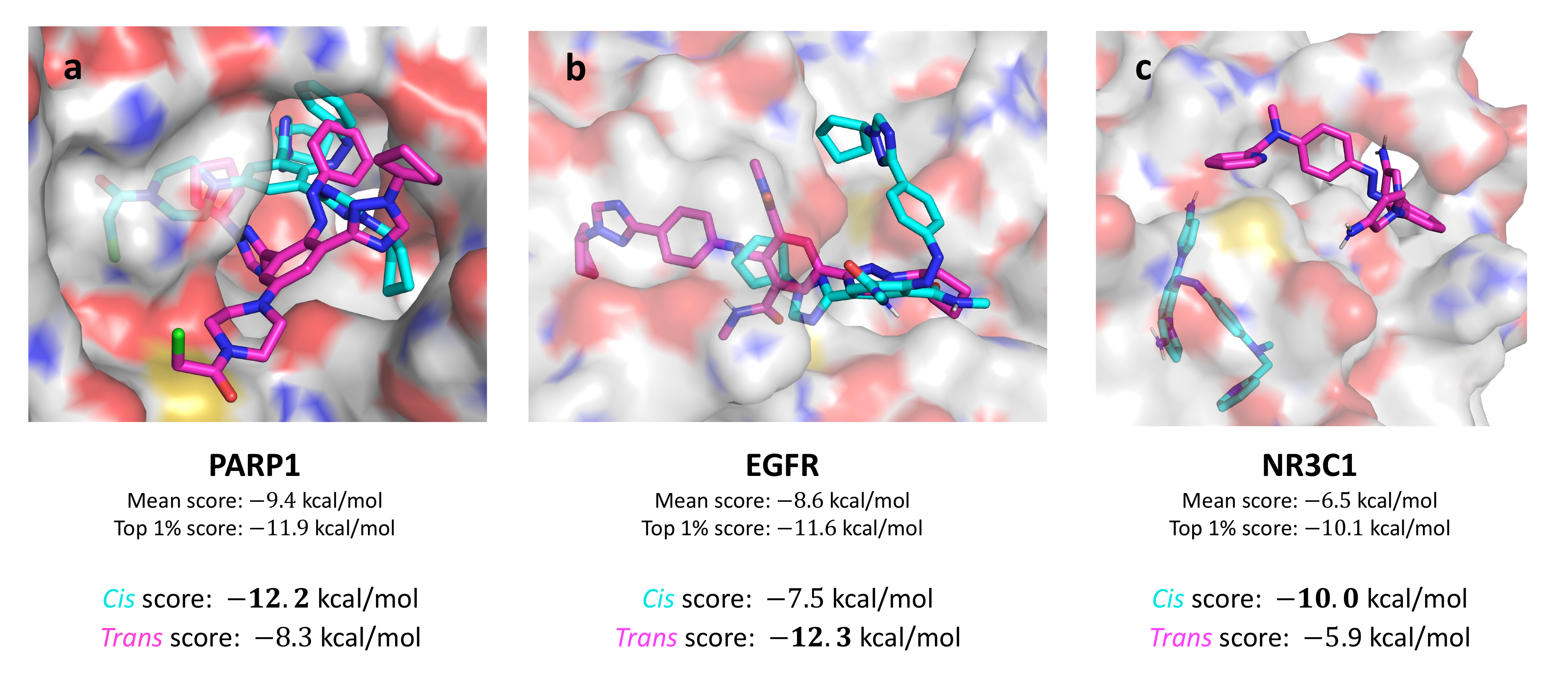}
    \caption{Binding poses of \textit{cis} and \textit{trans} isomers for ligands with high differential affinity. The isomer docking scores and the mean scores of all ligands are given below. The lower of the two isomers' scores are shown in bold.}
    \label{fig:docked_poses}
\end{figure*}


\begin{figure*}[t!]
    \centering
    \includegraphics[width=\textwidth]{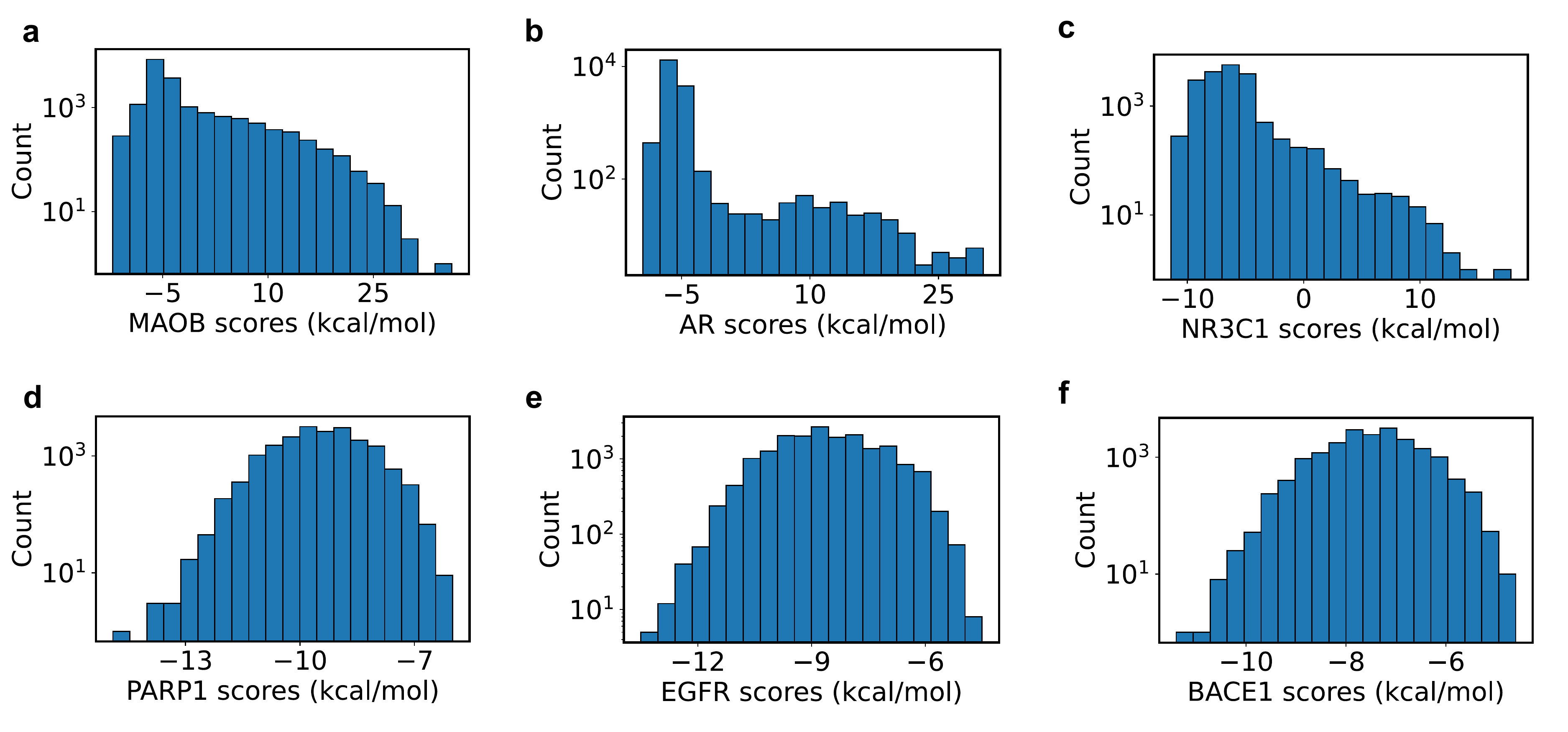}
    \caption{Distribution of docking scores for selected targets. Panels (a)-(c) show targets with wide ranges in score and differential affinity (see Fig. \ref{fig:cis_trans_correlation}(a) for the latter). Panels (d)-(f) show targets with typical ranges that are much smaller. Note the log scale of the $y$-axes.}
    \label{fig:outlier_scores}
\end{figure*}

\section*{Methods}
Ligands were selected from the 19,000
azobenzene derivatives generated in Ref. \cite{axelrod2022thermal}. They were created through combinatorial substitution of azobenzene with common literature substituents and patterns. We previously computed their absorption wavelengths and thermal half-lives using a machine learning (ML) force field trained on quantum chemistry data \cite{axelrod2022thermal}. We also argued that thermal isomerization proceeds through intersystem crossing for unsubstituted azobenzene. We therefore computed rates with both intersystem crossing and standard transition state (TS) theory for all derivatives. 

We performed ligand-protein docking using the recently developed \textsc{dockstring} package \cite{garcia2022dockstring}. \textsc{Dockstring} performs automated docking for 58 different proteins of medical interest, using only the ligand SMILES string as input. All proteins and docking search boxes were manually prepared by the \textsc{dockstring} authors. The program performs automatic ligand protonation and partial charge generation with Open Babel \cite{o2011open}, initial structure generation with RDKit \cite{rdkit}, and docking with AutoDock Vina \cite{trott2010autodock}. We added hydrogens to the exported poses with the \texttt{AddHs} function in RDKit, using the argument \texttt{addCoords=True}. We used RDKit version 2020.09.1 for all calculations.

\textsc{Dockstring}'s 58 proteins span a variety of functions: they contain 22 kinases, twelve enzymes, nine nuclear receptors, seven proteases, five G-protein coupled receptors, two cytochromes, and one chaperone \cite{garcia2022dockstring}. The accuracy of the \textsc{dockstring} procedure was tested through a comparison to experimental binding affinities from the ExCAPE database \cite{sun2017excape}. Results were of high quality for 24 targets, with enrichment factors between 4 and 6 for each target \cite{garcia2022dockstring}. High-quality targets discussed below include JAK2, PARP1, NR3C1, and HSD11B1. Medium-quality targets include EGFR, MAOB, AR, BACE1, MAPK1 and ADRB1. 

We removed all ligands that were protonated or de-protonated by Open Babel, since their chemical properties were computed without this step in Ref. \cite{axelrod2022thermal}. This left approximately 9,000 compounds in total. We docked both the \textit{cis} and \textit{trans} isomers for each compound for all 58 protein targets, yielding over one million docking scores in total. We removed all molecules with docked poses that did not respect the \textit{cis}/\textit{trans} isomerism of the input SMILES.

As discussed in the Results section, we validated ML predictions of photophysical properties for a number of hits. As in the ML training set \cite{axelrod2022thermal}, we used spin-flip time-dependent density functional theory (SF-TDDFT) with the BHHLYP functional \cite{becke1993new}, the 6-31G* basis \cite{francl1982self}, and the C-PCM implicit solvent model of water \cite{truong1995new, barone1998quantum, cossi2003energies}. Calculations were performed in Q-Chem 5.3 \cite{epifanovsky2021software}.

We also simulated the full absorption spectrum for these species. To do so we performed TDDFT with the PBE0 functional \cite{adamo1999toward} on geometries sampled from MD. PBE0 is among the most reliable functionals for TDDFT vertical excitation energies of organic molecules when compared to high-accuracy theoretical methods \cite{jacquemin2009extensive}. Further, as discussed in the Results section, it yields excellent agreement with the absorption spectrum of azobenzene when combined with MD. The range-separated hybrid CAM-B3LYP \cite{yanai2004new} has similar errors to PBE0 \cite{jacquemin2009extensive}, and has been used to simulate the spectrum of azobenzene \cite{gelabert2023predicting}, but we found that PBE0 gave slightly better agreement with experiment.

We used our ML force field from Ref. \cite{axelrod2022excited} to perform 300 ps of MD in the canonical ensemble at room temperature. We used the Nos\'e-Hoover thermostat \cite{nose1984unified, hoover1985canonical} with the parameters of Ref. \cite{axelrod2022excited}. We then sampled 2,000 frames from each simulation and performed TDDFT calculations on each. We used the large def2-TZVP basis set \cite{weigend2005balanced} and the SMD model for water \cite{marenich2009universal}. Calculations were performed with Orca 4.1.1 \cite{neese2012orca}. The spectrum was computed as the weighted histogram of excitation energies, with weights given by the square of the transition dipole moment, following Fermi's Golden Rule \cite{fermi1995notes}.

\section*{Results}
\subsection*{Differential affinity}
\subsubsection*{General trends}

Large-scale docking allows us to discover trends in photo-druggability. For example, we can learn whether the binding affinities of the two isomers tend to be correlated or not. Since the aim of photopharmacology is to develop drugs that are potent in one form and harmless in another, we can learn whether this goal is easy or hard. This, in turn, tells us about the size of the photoswitchable chemical space that contains promising photo-drug candidates.

Figure \ref{fig:cis_trans_correlation}(a) shows the distribution of \textit{cis} minus \textit{trans} docking scores for different proteins. Since a lower docking score means better binding, a positive value of $\Delta \equiv S_{\mathrm{c}} - S_{\mathrm{t}}$ means that \textit{trans} binds better than \textit{cis}. These differences are summarized in panel (b), which shows the Spearman rank correlation coefficients $\rho$ between \textit{cis} and \textit{trans} docking scores. Each data point is the correlation of isomer scores for a different target. The rank coefficient measures the extent to which the ordering of scores is correlated. We see that the scores are moderately to strongly correlated. The mean value of $\rho$ is 0.58; the correlation is above 0.5 for 66\% of targets, and above 0.75 for 21\% of them. Panel (c) shows the median absolute difference in \textit{cis}/\textit{trans} docking scores for the different targets. The majority of median differences are under 1 kcal/mol; the mean of medians is 0.8 kcal/mol, while the average standard deviation of scores is 1.2 kcal/mol. This again reflects the correlation of the \textit{cis}/\textit{trans} scores.

A more informative metric for score similarity is the enrichment factor (EF). The goal of virtual screening is to find the small proportion of ligands that tightly bind the target. Hence the Spearman correlation among ligands that do not tightly bind in either conformation, which is most of them, is not of much interest. The EF is more related to the problem of virtual screening. It is defined as the ratio of true hit percentage in a model's top-ranked subset to the hit percentage in the whole dataset. Here we ask what the enrichment factor is from using \textit{cis} scores to predict \textit{trans} scores. Defining a hit as a molecule in the lowest (best) 1\% of all \textit{trans} scores, the EF is then the true hit percentage in the lowest 1\% of \textit{cis} scores, divided by the true hit percentage in the dataset (equal to 1). We see in panel (d) that enrichment factors are quite high for most targets, with a mean value of 23.6. Hence the likelihood of a \textit{trans} hit also being a \textit{cis} hit is rather high. This means that finding a molecule with a large score difference is a needle-in-a-haystack problem, rather than a trivial one.

Thus we are able to map the space of photodruggable targets and photoactive ligands for the first time, and find that most isomers are too similar to be good candidates for photopharmacology. Indeed, taking the docking score as an estimate of the binding free energy, we require a score difference of 4.1 kcal/mol to achieve a 1000-fold difference in dissociation constant $K_{\mathrm{d}}$. This is 5.1 times higher than the average difference in \textit{cis} and \textit{trans} scores (Fig. \ref{fig:cis_trans_correlation}(c)). Only 0.79\% of all ligand-target pairs have \textit{cis}/\textit{trans} score differences above 4.1 kcal/mol. Of the ligands in the top $5^{\mathrm{th}}$ percentile of all scores, only 4\% have the requisite score difference. This finding is significant because it suggests that most azobenzene-based inhibitors do not have strong differential activity. However, this prediction should be taken with caution, since docking has several well-known limitations that can affect accuracy \cite{du2016insights}. 

\subsubsection*{Target-specific trends}
We also find that some targets are more easily photodruggable than others. In Fig. \ref{fig:cis_trans_correlation}(a), we see that $\mathrm{mean}\{ \Delta \}$ is positive for most targets, and hence that \textit{trans} tends to bind better on average. The mean value of $\Delta$ over all targets is $0.53$ kcal/mol. The targets with a large value of $\Delta$ are more easily photodruggable, because \textit{trans} ligands with strong affinities are more likely to have \textit{cis} counterparts with weak affinities.

There are several targets for which \textit{trans} is particularly favored over \textit{cis}. For example, the mean value of $\Delta$ is over 1.5 kcal/mol for EGFR, JAK2, MAPK1 and ADRB1. EGFR shows the strongest difference in scores, with $\mathrm{mean} \{ \Delta \} = 1.7$ kcal/mol. For context, this is the difference between the mean and the $6^{\mathrm{th}}$ percentile EGFR scores in the \textsc{dockstring} dataset. It therefore represents a significant difference.  The high differential affinity in EGFR comes from both high \textit{trans} affinity and low \textit{cis} affinity. For example, the mean EGFR  score is $-8.9$ kcal/mol in the \textsc{dockstring} dataset, while the mean \textit{cis} and \textit{trans} scores in our library are $-7.8$ and $-9.5$ kcal/mol, respectively. 

We can understand these differences by visualizing the docked poses. The poses of ligands docked in three different targets are shown in Fig. \ref{fig:docked_poses}. Panel (b) shows the ligands with the highest differential activity in EGFR. We see that the \textit{trans} isomer fills a long and narrow binding pocket. The \textit{cis} isomer is kinked, giving it two arms that extend only half the length of the pocket. Hence the \textit{trans} isomer binds much more strongly than \textit{cis}. The shape of the binding pocket thus explains the strong preference for \textit{trans} ligands in EGFR.

While less common, there are also some targets with a preference for \textit{cis} isomers. For example, the mean value of $\Delta$ is less than $-0.3$ kcal/mol for HSD11B1, MAOB and PARP1. The lowest value is $-0.44$ kcal/mol for HSD11B1, which should be contrasted with the highest value of $1.7$ kcal/mol for EGFR. Figure \ref{fig:docked_poses}(a) shows the tight four-pronged fit of a \textit{cis} ligand in PARP1, and the lack of space for the \textit{trans} isomer. 

\begin{figure*}[t!]
    \centering
    \includegraphics[width=\textwidth]{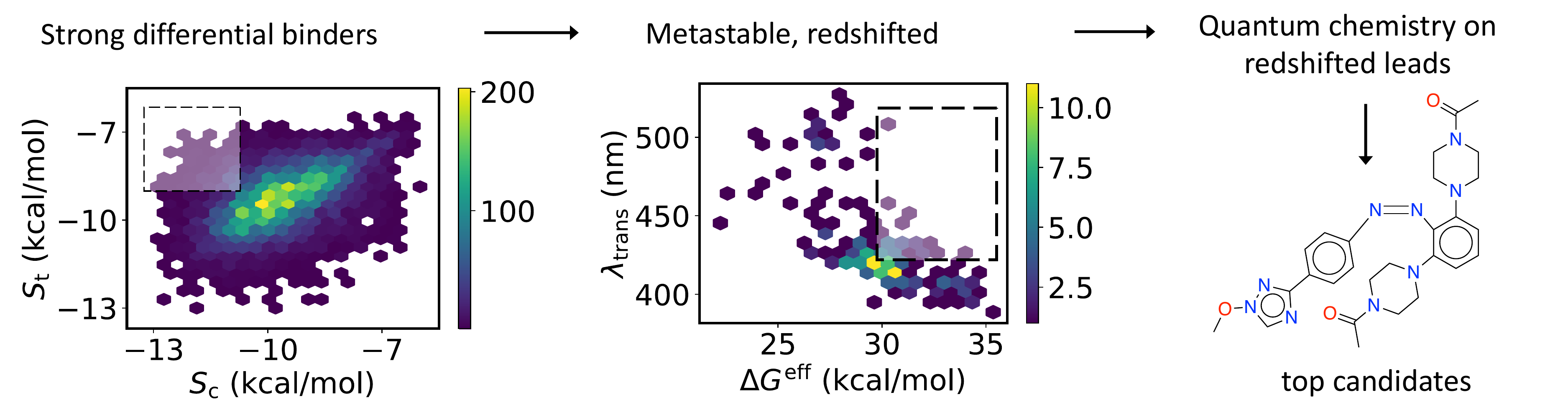}
    \caption{Workflow for virtual screening of photoactive drugs. The example given is for the PARP1 target. From left to right, we first screen for ligands that are strong binders in \textit{cis} form and poor binders in \textit{trans} form. We then narrow down this subset to molecules with high thermal isomerization barriers and redshifted  absorption wavelengths. Lastly, we perform quantum chemistry calculations on the hits with the highest absorption wavelengths, and then select the top candidates. }
    \label{fig:screening_workflow}
\end{figure*}

Since some targets have a strong preference for \textit{trans}, but only a few have a moderate preference for \textit{cis}, it is easier to design \textit{trans}-active drugs in general. This is a significant finding for photopharmacology, since it informs the community about the ease of designing active drugs in different isomeric forms.  It is also not ideal, since one usually wants to deliver an inactive drug and activate it locally, and it is easier to deliver the thermodynamically more stable \textit{trans} isomer. (\textit{Trans} is more stable than \textit{cis} in over 99\% of the molecules in our dataset \cite{axelrod2022thermal}.) Hence one usually wants the \textit{trans} isomer to be inactive and the \textit{cis} isomer to be active. Docking of bridged azobenzenes, in which \textit{cis} is more stable than \textit{trans} \cite{siewertsen2009highly}, may therefore be of interest in the future. Nevertheless, one can still find ligand-target pairs in which \textit{cis} binds much more strongly than \textit{trans}, as shown in Figs. \ref{fig:docked_poses} and \ref{fig:screening_workflow} and discussed further below.

Some targets have a very wide range of score differences, which makes them particularly amenable to photopharmacology. For example, the score differences for MAOB, AR, and NR3C1 extend past the plot limits in Fig. \ref{fig:cis_trans_correlation}(a). Indeed, the standard deviations in \textit{cis}-\textit{trans} score differences are 6.6 kcal/mol for MAOB, 3.4 kcal/mol for AR, and 2.4 kcal/mol for NR3C1, compared to only 0.7 kcal/mol for PARP1.
The $5^{\mathrm{th}}$ and $95^{\mathrm{th}}$ percentile of score differences for MAOB are $-13.1$ and $10.5$ kcal/mol, respectively, which represent an enormous range.

Figure \ref{fig:outlier_scores} shows docking scores for targets with wide and normal ranges of differential affinity. We see that many ligands docked in MAOB, AR, and NR3C1 have positive scores, while those docked in PARP1, EGFR and BACE1 do not. Hence for many ligands, it is thermodynamically unfavorable to be bound to MAOB, AR or NR3C1 at all. This can also be seen in the original \textsc{dockstring} dataset. The reason is that the binding pockets of these proteins are quite small, meaning that many ligands have clashes with the target. Indeed, Fig. \ref{fig:docked_poses}(c) shows a \textit{trans} ligand that has a better score when docked outside the pocket than within it. Hence these targets are difficult to drug, which makes them good candidates for photopharmacology. 

\subsection*{Virtual screening}
With access to docking scores and photophysical properties computed in Ref. \cite{axelrod2022thermal}, we can also perform virtual screening to identify photo-drug hits. Figure \ref{fig:screening_workflow} shows our screening workflow. We first identify ligands with strong \textit{cis} binding affinities and weak \textit{trans} affinities. This is the ideal case for photopharmacology, but, as mentioned above, is generally difficult because of the correlation between \textit{cis} and \textit{trans} scores, and the preference of most proteins for \textit{trans} isomers. Here we retain compounds for which $S_{\mathrm{c}} - \mu < -1.5 \mathspace \sigma$ and $S_{\mathrm{t}} - \mu > -0.5 \mathspace \sigma$, where $\mu$ is the mean docking score of all ligands for the given target, and $\sigma$ is the associated standard deviation. More extreme cutoffs can be used when screening larger ligand libraries. 

From these compounds, we retain only species with longer \textit{cis} thermal half-lives than unsubstitued azobenzene. The experimental half-life of azobenzene is 1.4 days in benzene solution at $35^{\circ}$ C \cite{talaty1967thermal}; hence this cutoff should select for lifetimes that are at least several hours. We keep all compounds with $\Delta G^{\mathrm{eff}} \geq \Delta G^{\mathrm{eff}}_{\mathrm{azobenzene}}$. We use the effective activation free energy, $\Delta G^{\mathrm{eff}}$, which measures the barrier to triplet-mediated thermal isomerization \cite{axelrod2022thermal}, and which is analogous to $\Delta G^{\dagger}$ in Eyring transition state theory. 
 Lastly, we select the remaining compounds with the highest \textit{trans} absorption wavelengths, and validate their photophysical properties with quantum chemistry. 

\begin{figure*}[t!]
    \centering
    \includegraphics[width=\textwidth]{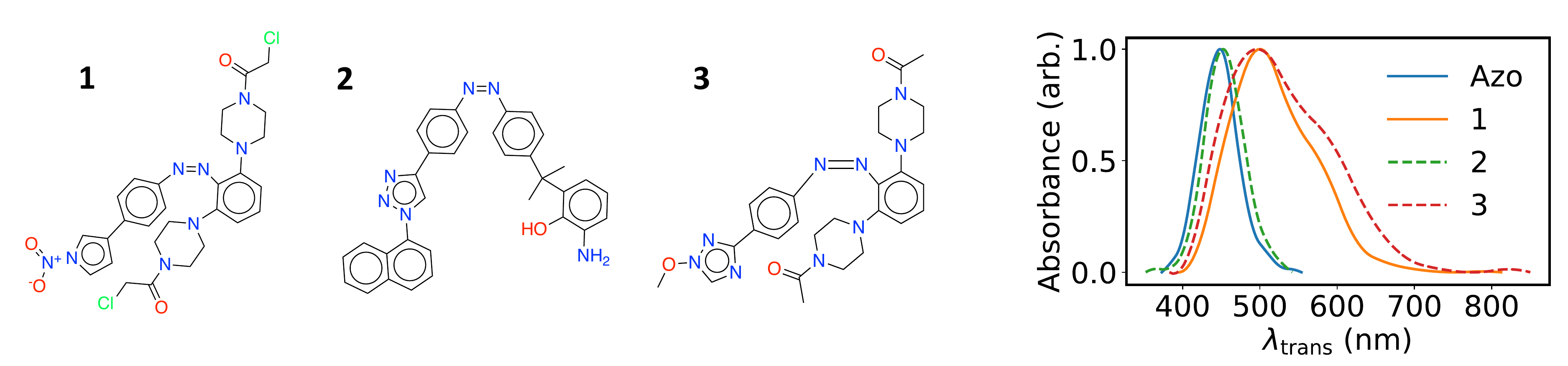}
    \caption{Bioactive \textit{cis} isomers of hits \textbf{1}-\textbf{3} (left), and \textit{trans} absorption spectra of both the hits and unsubstituted azobenzene (right). The spectra of \textbf{1} and \textbf{3} extend into the therapeutic window of 700-900 nm.}
    \label{fig:parp1_hits_and_spectrum}
\end{figure*}

Table \ref{tab:parp1_hits} shows docking scores and chemical properties for three promising photoactive inhibitors of PARP1. For reference, the second last row shows results for azobenzene, and the last row shows median values of all molecules. The promising candidates are shown in Fig. \ref{fig:parp1_hits_and_spectrum}. The \textit{cis} docking scores of the hits are all 2 kcal/mol or more below the median. The \textit{trans} docking scores are no more than $0.3$ kcal/mol below the median. The score differences are all at least 2.1 kcal/mol. The largest difference is for compound \textbf{2}, which also has the lowest \textit{cis} docking score (0.4 percentile among all compounds). 
\begin{table*}[t!]
\centering
\begin{tabular}{c||c|c||c|c|c||c|c||c|c}
     \hline
     Compound &  $  S^{\mathrm{dock}}_{\mathrm{cis}} $  &  $S^{\mathrm{dock}}_{\mathrm{trans}} $   & $\lambda_{\mathrm{trans}}^{\mathrm{ML}}$ &  $\lambda_{\mathrm{trans}}^{\mathrm{SF}}$ & $\lambda_{\mathrm{trans}}^{\mathrm{PBE0}}$ &  $\Delta G^{\mathrm{eff}}_{\mathrm{ML}}$ &  $\Delta G^{\mathrm{eff} }_{\mathrm{SF}}$ &  $\Delta G^{\dagger}_{\mathrm{ML}}$ &  $\Delta G^{\dagger}_{\mathrm{SF}}$ \\ \hline
     \textbf{1} & $-11.4$ & $-9.0$ & 511 & 420 & 518 &  30.2 & -- & 31.2 & 31.9  \\ 
     \textbf{2} & $-12.3$ & $-9.7$ & 502 & -- & 482 & 29.2 & 30.0 & 30.2 & 30.9 \\ 
     \textbf{3} & $-$11.6 & $-$9.5 & 456 & 467 & 532 & 30.0 & -- & 31.0 & 31.8 \\  \hline
     Azobenzene &  $-6.0$ & $-6.7$ & 410 & 411 & 457 & 28.9 & 28.8 & 30.3 & 30.2 \\ \hline
     \begin{tabular} {@{}c@{}}Median, all \vspace*{-0.1cm} \\  molecules \end{tabular}
     & \multicolumn{2}{c||}{$-9.4$}  & 426 & -- & -- &  29.5 & -- & 29.5  & --  \\ 
    \hline
\end{tabular}
\caption{Properties of three hits for photoactive inhibition of PARP1. For reference we also give the properties of azobenzene, together with the median among all molecules. From left to right: \textit{cis} and \textit{trans} docking scores (kcal/mol); absorption wavelength of the \textit{trans} isomer, computed with ML, SF-TDDFT with the BHHLYP functional, and TDDFT with the PBE0 functional (nm); triplet-mediated effective activation free energy for thermal \textit{cis}-\textit{trans} isomerization, computed with ML and SF-TDDFT (kcal/mol); and Eyring activation free energy, computed with ML and SF-TDDFT (kcal/mol). ML and SF-TDDFT calculations were performed on ML-optimized geometries, while TDDFT with PBE0 was performed on PBE0-optimized geometries. Missing SF-TDDFT properties for compounds \textbf{1}-\textbf{3} could not be reliably determined due to spin contamination. }
\label{tab:parp1_hits}
\end{table*}

ML calculations predict that compounds \textbf{1} and \textbf{2} have \textit{trans} absorption wavelengths 
($\lambda_{\mathrm{trans}} $) over 500 nm. This is a redshift of over 90 nm relative to azobenzene. The predicted absorption wavelength of compound \textbf{3} is only 456 nm. To test these predictions, we performed quantum chemistry calculations with SF-TDDFT using the BHHLYP functional and an implicit model of water, which was the method used for training. Interestingly, these calculations yielded $\lambda_{\mathrm{trans}} = 420$ nm for \textbf{1} and $\lambda_{\mathrm{trans}} = 467$ nm for \textbf{3}. Hence the extreme redshift predicted for \textbf{1} was significantly overestimated by the ML model. Reliable results could not be obtained for \textbf{2} due to spin contamination in SF-TDDFT. Both spin contamination and overestimated redshifting in high-$\lambda$ species were previously noted in Ref. \cite{axelrod2022thermal}.

SF-TDDFT was used in Ref. \cite{axelrod2022thermal} because of its ability to handle multireference effects in azobenezene transition states. However, it has not been thoroughly benchmarked for absorption wavelengths, and its predicted $n-\pi^*$ absorption maximum for azobenzene is 30 nm lower than experiment \cite{axelrod2022excited}. Further, vertical excitation energies with any functional do not account for the redshift \cite{gelabert2023predicting} or blueshift \cite{axelrod2022excited} that can occur at nonzero temperatures. They also do not account for the width of the absorption spectrum. Since one is usually interested in the high-wavelength tail of the spectrum, predicting its width is critical.

Therefore, we also computed the full absorption spectrum of each hit, using MD combined with TDDFT with the PBE0 functional (see Methods). The computed absorption spectrum for each hit is shown in Fig. \ref{fig:parp1_hits_and_spectrum}. Both \textbf{1} and \textbf{3} are predicted to absorb in the therapeutic window of 700-900 nm \cite{dong2017near}, which is a major goal for photopharmacology. The spectrum of azobenzene is also shown for reference. It is in excellent agreement with experiment: the predicted peak wavelength for azobenzene is 448 nm, and the absorption extends to 553 nm; the experimental maximum is 444 nm in acetonitrile, and the extent is also around 550 nm \cite{knie2014ortho}. This validates our approach and provides some confidence in the spectra of \textbf{1}-\textbf{3}.

The predicted near-IR absorption comes from a significantly widened, redshift-skewed spectrum and a redshifted maximum. The maxima of \textbf{1} and \textbf{3} are 501 and 496 nm, which are 53 and 48 nm higher than azobenzene, respectively. However, if these compounds had the same spectrum shape as azobenzene, then their maximum absorption would only extend to 600 nm. It is the long tail of redshifted absorption that enables near-IR absorption. The large \textit{ortho} groups lead to twisting around the central nitrogen double bond, and hence to sampling of low-gap geometries. We reported similar effects from bulky substitution in Ref. \cite{axelrod2022excited}. Redshift-skewed absorption has been measured experimentally in tetra-\textit{ortho} halogenated azobenzenes \cite{konrad2020computational}, and reproduced with MD simulations in Ref. \cite{gelabert2023predicting}.

For comparison, we also re-optimized the equilibrium \textit{trans} geometries with PBE0 and D3 dispersion \cite{grimme2010consistent}, and then performed single-point PBE0 TDDFT calculations. The results are shown in Table \ref{tab:parp1_hits} under the heading $\lambda_{\mathrm{trans}}^{\mathrm{PBE0}}$. Some of the vertical excitation wavelengths match the MD means, but all are blueshifted with respect to the MD maxima. This reinforces the need for a full MD simulation of the spectrum. 

\begin{figure*}[t!]
    \centering
    \includegraphics[width=\textwidth]{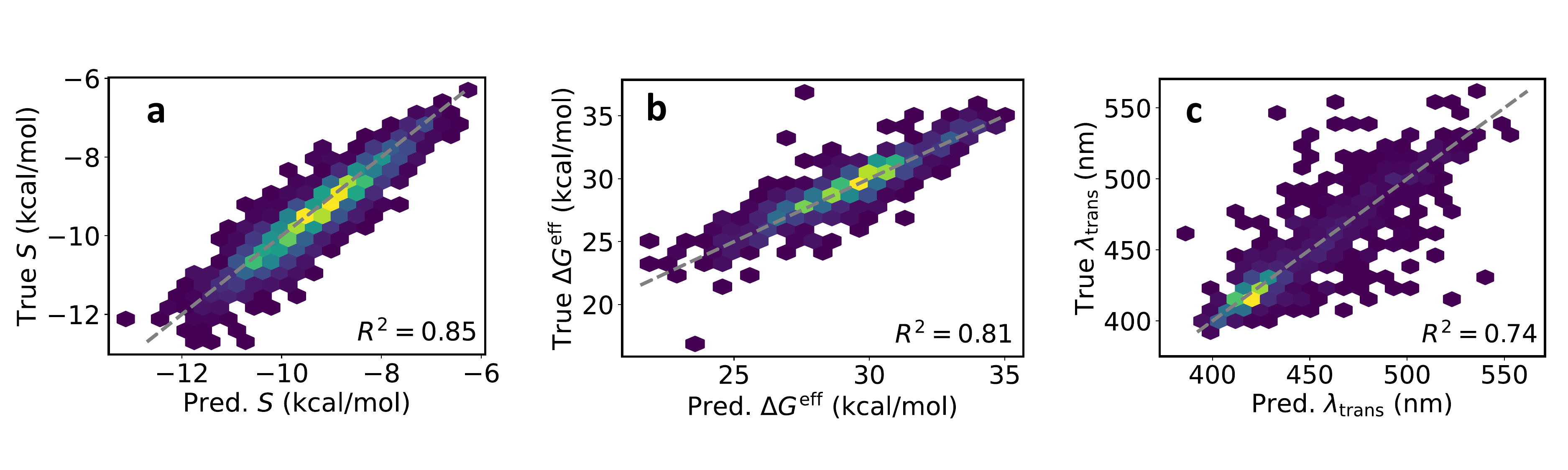}
    \caption{Test set accuracy of graph-to-property models. Perfect correlation is shown with a dashed gray line. (a) Docking score for PARP1. (b) Effective activation free energy for thermal isomerization. (c) \textit{Trans} absorption wavelength. }
    \label{fig:graph_models}
\end{figure*}
Each hit also has a longer predicted thermal half-life than azobenzene. $\Delta G^{\mathrm{eff}}_{\mathrm{ML}}$ is 28.9 kcal/mol for azobenzene \cite{axelrod2022thermal}, and over 29.2 kcal/mol for compounds \textbf{1}-\textbf{3}. Note that the experimental value for azobenzene is 25.4 kcal/mol \cite{rietze2017thermal}, and that the overestimate comes from systematic errors in SF-TDDFT, though experimental trends are well-reproduced \cite{axelrod2022thermal}. We also computed $\Delta G^{\dagger}$, the activation free energy from Eyring TS theory, since isomerization can occur through a singlet TS if $\Delta G^{\dagger} < \Delta G^{\mathrm{eff}}$. The TS barrier of each hit is comparable to or higher than that of azobenzene.

We further performed quantum chemistry calculations to estimate $\Delta G^{\mathrm{eff}}_{\mathrm{SF}}$ and $\Delta G^{\dagger}_{\mathrm{SF}}$, the free energies from SF-TDDFT. In particular, we performed single-point calculations on the ML-optimized singlet-triplet crossing geometries and TSs. We then used the ML estimates of the remaining contributions to $\Delta G$ to approximate $\Delta G^{\mathrm{eff}}_{\mathrm{SF}}$ and $\Delta G^{\dagger}_{\mathrm{SF}}$. Spin contamination led to unreliable estimates for $\Delta G^{\mathrm{eff}}$ in \textbf{1} and \textbf{3}. However, the reliable estimate for \textbf{2} and all estimates for $\Delta G^{\dagger}$ were within 0.8 kcal/mol of the predicted values. Hence the quantum chemical calculations also indicate long thermal half-lives for these species.

Interestingly, our screen produced two very similar molecules in \textbf{1} and \textbf{3}. The ortho functional groups are identical, except that chlorine is removed in \textbf{3}, yielding a ketone instead of a chloroacetate. The nitro-substituted pyrrole in \textbf{1} is replaced with a methoxy-substituted triazole in the same position. The compounds have similar absorption spectra, thermal isomerization barriers, and differential affinities. Compound \textbf{3} is preferrable to \textbf{1}, since chloroacetates are hazardous alkylating agents that can be toxic to cells, and nitro groups can be toxic as well. Removing chlorine and replacing the nitro group with a methoxy group thus makes \textbf{3} a more promising candidate. Automatic pre-filtering to remove problematic motifs is of interest for future work. 

\subsection*{Accelerated screening with graph-based models}

Using ML potentials to compute thermal half-lives and absorption wavelengths is significantly faster than using quantum chemistry. However, the calculations are still rather slow, because they involve multiple conformer searches and optimizations, which take several minutes for each compound. Hence only 25,000 molecules could be screened in Ref. \cite{axelrod2022thermal}. This falls well short of the millions of compounds routinely screened in computational drug discovery \cite{zhang2021development}. Therefore, we propose accelerating future virtual screening efforts with graph-to-property models trained on ML-accelerated simulations. These models are quite fast, and can therefore be used to screen hundreds of millions of compounds for a desired property \cite{stokes2020deep}. Further, while ML-accelerated calculations are too slow for such large-scale screening, they are fast enough to generate tens of thousands of training points to train high-accuracy graph-based models. Such models can also be trained on docking scores, yielding smaller but still significant savings.

Graph-to-property models take the chemical graph as input and yield a property as output. Unlike models that take one or many 3D structures as input \cite{axelrod2020molecular, 2020geom}, graph-based models do not require expensive generation and refinement of 3D coordinates. They are therefore ideal for rapid virtual screening of large libraries. 

Among the many models developed in the literature, the Attentive FP architecture is one of the most successful \cite{xiong2019pushing}. It combines an attention mechanism \cite{noam_lr} with neural network message-passing \cite{gilmer2017neural} to generate an embedding for each atom. These node embeddings are then converted to graph embeddings with attention-based pooling, before being mapped to the final property with a neural network. This model substantially improves on previous state-of-the-art results for nearly all properties in standard benchmarks \cite{xiong2019pushing}. Further, it has the highest accuracy of all models trained on the \textsc{dockstring} dataset for each protein target \cite{garcia2022dockstring}. We therefore used the Attentive FP model in this work.

We used the DeepChem library \cite{Ramsundar-et-al-2019} to train separate Attentive FP models on PARP1 docking scores, effective activation free energies, and \textit{trans} absorption wavelengths. We trained each model for 500 epochs using a batch size of 16 and a learning rate of 0.001. Absorption wavelengths were first standardized to have zero mean and unit variance, which accelerated training. 10\% of the compounds were randomly selected as a test set. 

The results are shown in Fig. \ref{fig:graph_models}. We see that all models are highly predictive, and hence can be used to accelerate virtual screening. The models for $S$ and $\Delta G^{\mathrm{eff}}$ are particularly good, with coefficients of determination $R^2 > 0.8$. The mean absolute error (MAE) of the docking model is 0.31 kcal/mol, and the enrichment factor is 44 when defining a hit as the lowest 1\% of all scores. The MAE is 0.68 kcal/mol for $\Delta G^{\mathrm{eff}}$ and 9.9 nm for $\lambda_{\mathrm{trans}}$. $R^2$ is only 0.74 for $\lambda_{\mathrm{trans}}$, and more outliers are clearly visible for this model than for the others. This may be because the ML force field, which generated data for the graph model, is less accurate for $\lambda_{\mathrm{trans}}$ than for $\Delta G^{\mathrm{eff}}$ \cite{axelrod2022thermal}. This may lead to artificial activity cliffs that are hard for the graph model to predict. Nevertheless, the model is still accurate enough to be useful in virtual screening.

\section*{Conclusion}
We have used computational docking to predict the \textit{cis} and \textit{trans} binding affinities of 9,000 azobenzene derivatives for 58 different proteins. This revealed substantial differences in the photo-druggability of different targets. We then combined these docking scores with previous calculations of the absorption wavelength and thermal stability to identify photoactive hits for the PARP1 cancer target. Lastly, we showed that this data enables training of fast graph-to-property models that can significantly accelerate future virtual screening efforts. 

Our virtual screening results show that docking and ML can be used to identify promising photoactive drug candidates. Our large-scale docking results also yield important insights for photopharmacology. First, most proteins tend to prefer \textit{trans} isomers over \textit{cis}, which is problematic if one wants to deliver a thermodynamically stable and inactive compound. Second, the binding affinities of \textit{cis} and \textit{trans} ligands tend to be highly correlated. This means that even if one designs a potent drug based on azobenzene, it is likely that the other isomer is also potent. Third, while these trends are discouraging for photopharmacology, they do not apply to every protein and every ligand. Indeed, some proteins prefer \textit{cis} isomers over \textit{trans}, and for each protein there are ligand outliers that have strong differential activity. Further, proteins with small binding pockets, such as MAOB and NR3C1, show large differences in isomeric affinity. These insights will be crucial for the future design of photoactive drugs.

\section*{Acknowledgements} \label{sec:acknowledgements}
We thank Professor Zlatko Janeba (IOCB Prague) for enlightening discussions. Harvard Cannon cluster, MIT Engaging cluster, and MIT Lincoln Lab Supercloud cluster at MGHPCC are gratefully acknowledged for computational
resources and support. Financial support from DARPA (Award HR00111920025) is acknowledged.





\bibliography{main}

\end{document}